\documentclass[12pt]{article}

\usepackage{amsmath,amssymb,amsthm,amscd,pstricks}
\usepackage{graphicx,tocloft}
\usepackage[nosort]{cite}
\def\figurename{Figure}

\makeatletter

\renewcommand{\fnum@figure}[1]{\textbf{\figurename~\thefigure}:}

\makeatother

\makeatletter

\renewcommand\section{\@startsection{section}{1}{\z@}
                                   {-3.5ex \@plus -1ex \@minus -.2ex}
                                   {2.3ex \@plus .2ex}
                                   {\normalfont\large\bfseries}}
\renewcommand\subsection{\@startsection{subsection}{2}{\z@} 
                                   {-3.25ex\@plus -1ex \@minus -.2ex}
                                   {1.5ex \@plus .2ex}
                                   {\normalfont\normalsize\bfseries}}
\renewcommand\subsubsection{\@startsection{subsubsection}{3}{\z@}
                                   {-3.25ex\@plus -1ex \@minus -.2ex}
                                   {1.5ex \@plus .2ex}
                                   {\normalfont\normalsize\bfseries}}
\renewcommand\paragraph{\@startsection{paragraph}{4}{\z@}
                                   {3.25ex \@plus1ex \@minus.2ex}
                                   {-1em}
                                   {\normalfont\normalsize\bfseries}}

\makeatother

\newdimen\tableauside\tableauside=1.0ex
\newdimen\tableaurule\tableaurule=0.4pt
\newdimen\tableaustep
\def\phantomhrule#1{\haox{\vbox to0pt{\hrule height\tableaurule
width#1\vss}}}
\def\phantomvrule#1{\vbox{\haox to0pt{\vrule width\tableaurule
height#1\hss}}}
\def\sqr{\vbox{%
  \phantomhrule\tableaustep

\haox{\phantomvrule\tableaustep\kern\tableaustep\phantomvrule\tableaustep}%
  \haox{\vbox{\phantomhrule\tableauside}\kern-\tableaurule}}}
\def\squares#1{\haox{\count0=#1\noindent\loop\sqr
  \advance\count0 by-1 \ifnum\count0>0\repeat}}
\def\tableau#1{\vcenter{\offinterlineskip
  \tableaustep=\tableauside\advance\tableaustep by-\tableaurule
  \kern\normallineskip\haox
    {\kern\normallineskip\vbox
      {\gettableau#1 0 }%
     \kern\normallineskip\kern\tableaurule}%
  \kern\normallineskip\kern\tableaurule}}
\def\gettableau#1 {\ifnum#1=0\let\next=\null\else
  \squares{#1}\let\next=\gettableau\fi\next}

\newcommand{\be}{\begin{equation}}
\newcommand{\ee}{\end{equation}}
\newcommand{\bea}{\begin{eqnarray}}
\newcommand{\eea}{\end{eqnarray}}
\newcommand{\ba}{\begin{array}}
\newcommand{\ea}{\end{array}}
\newcommand{\id}{\haox{1\kern-.27em l}}
\newcommand{\oddots}{ \begin{picture}(12,12) \put(2,2){\circle*{1.3}}\put(7,6){\circle*{1.3}} \put(11,10){\circle*{1.3}}  \end{picture} }
\newcommand{\ZZ}{\mathbb{Z}}

\newcommand{\RR}{\mathbb{R}}
\newcommand{\half}{ {\textstyle \frac{1}{2}  } }

\newcommand{\de}{\delta}

\newcommand{\ep}{\epsilon}

\newcommand{\la}{\lambda}

\newcommand{\om}{\omega}

\newcommand{\La}{\Lambda}

\newcommand{\tha}{\theta}

\newcommand{\cN}{\mathcal{N}}

\newcommand{\cW}{\mathcal{W}}
\newcommand{\cM}{\mathcal{M}}

\newcommand{\rar}{\rightarrow}

\newcommand{\non}{\nonumber}
\newcommand{\lb}{\langle}
\newcommand{\rb}{\rangle}

\newcommand{\tia}{\tilde{a}}
\newcommand{\tI}{\tilde{I} }
\newcommand{\tell}{\tilde{\ell}}
\newcommand{\tY}{\tilde{Y}}

\newcommand{\SU}{\mathrm{SU}}

\newcommand{\SL}{\mathrm{SL}}

\newcommand{\sll}{\mathrm{sl}}

\newcommand{\ts}{\textstyle}

\begin{document}

\begin{center}
\vspace*{-2mm}
{\Large\sf
{Instanton partition functions in {\large $\cN=2$} {\large $\SU(N)$} gauge theories \\ with a general surface operator, and their $\cW$-algebra duals }}

\vspace*{6mm}
{\large Niclas Wyllard}

\vspace*{4mm}

{\tt n.wyllard@gmail.com}

\vspace*{8mm}
{\bf Abstract} 
\end{center}
\vspace*{0mm}
\noindent  
We write down an explicit conjecture for the instanton partition functions in $4d$ $\cN=2$ $\SU(N)$ gauge theories in the presence of a certain type of surface operator. These surface operators   are classified by partitions of $N$, and for each partition there is an associated partition function.  For the partition  $N\!=\!N$ we recover the Nekrasov  formalism, and when $N\!=\!1{+}\ldots{+}1$ we reproduce the result of Feigin et.~al. For the case $N\!=\!1+ (N{-}1)$ our expression is consistent with an alternative formulation in terms of a restricted $\SU(N){\times}\SU(N)$ instanton partition function. When $N\!=\!1{+}\ldots{+}1{+}2$ the partition functions can also be obtained perturbatively from certain $\cW$-algebras known as quasi-superconformal algebras, in agreement with a recent general proposal.

\vspace{1mm}

\setcounter{tocdepth}{1}

\setcounter{equation}{0}
\section{Introduction}\label{sint}

In \cite{Wyllard:2010} we argued that there is a general connection between instanton partition functions in $\cN=2$ gauge theories with a certain type of surface operator and a class of $\cW$-algebras (see also the earlier work \cite{Braverman:2010} which contains similar ideas). For the $\SU(N)$ gauge theories both the surface operators and the $\cW$-algebras are classified by partitions of $N$.

Surface operators in gauge theories are objects localised on two-dimensional submanifolds and the precise type that is relevant here is conveniently described using the $6d$ $(0,2)$ theory formulated on $\RR^4{\times}C$ \cite{Gaiotto:2009a,Witten:1995}, where  the $4d$ $\cN=2$ gauge theory lives on $\RR^4$ and a $2d$ conformal field theory lives on the Riemann surface $C$. In this language the surface operator arises from a $4d$ defect spanning a $2d$ submanifold of $\RR^4$ and wrapping $C$ \cite{Alday:2010}. For the $\SU(N)$ theories such surface operators are classified by partitions of $N$, since the $4d$ defects of the $6d$ $A_{N-1}$ $(0,2)$ theory have this classification \cite{Gaiotto:2009a}. 

The class of $\cW$-algebras that is relevant arises from affine Lie algebras via so called quantum  Drinfeld-Sokolov reduction \cite{Fateev:1987,Polyakov:1989,deBoer:1993}. For the case of the $\SU(N)$ gauge theories  the pertinent $\cW$-algebras appear by reduction from the affine $\sll_N$ algebra, and are classified by partitions of $N$  \cite{Bais:1990}. In the $6d$ language used above, the $\cW$-algebra may be thought of as the symmetry algebra of the $2d$ conformal field theory living on $C$. This harmonises nicely with the expectation that wrapping different $4d$ defects  on $C$ should change the $2d$ CFT (and in particular its symmetry algebra).

The proposal in \cite{Wyllard:2010} is a natural generalisation of  the various relations between $2d$ conformal field theories  and $4d$ $\cN\,{=}\,2$ gauge theories (with surface operators),  that have been discovered in the last year and a half. 

In particular, for the $A_1$ AGT relation \cite{Alday:2009a} (or its non-conformal version \cite{Gaiotto:2009b})  the relevant $\cW$-algebra is  the Virasoro algebra, and the instanton partition functions are those of the $\cN\,{=}\,2$ $\SU(2)$ gauge theories without a surface operator (the absence of a surface operator is conveniently thought of as a trivial surface operator). More generally for the $A_{N-1}$ AGT relation \cite{Wyllard:2009} (or its non-conformal version \cite{Taki:2009}), the relevant $\cW$-algebras are the $\cW_N$ algebras, and the instanton partition functions are those of the $\cN\,{=}\,2$ $\SU(N)$ gauge theories without a surface operator. For these cases the  instanton partition functions can be computed using the results of Nekrasov \cite{Nekrasov:2002}\footnote{More precisely, the Nekrasov formalism in its present form can only be used for  conventional gauge theories, and not for the generalized quivers \cite{Gaiotto:2009a}. In this paper we only discuss conventional theories.}.

Another class of examples correspond to $N=1{+}\ldots{+}1$ in the partition language. For these theories the relevant $\cW$-algebras are the $\widehat{\sll}_N$ (affine $\sll_N$) algebras. Relations of this type were first discovered in the non-conformal case in the mathematical literature several years ago  \cite{Braverman:2004a} (albeit using a different language).  For the conformal theories,  the relation involving $\widehat{\sll}_2$ was found in \cite{Alday:2010} and further studied in \cite{Awata:2010}. The extension to $\widehat{\sll}_N$ was treated in \cite{Kozcaz:2010b}.  For these cases the instanton partition functions can be computed using the results in \cite{Feigin:2008} (cf.~\cite{Alday:2010,Awata:2010,Kozcaz:2010b}). 

To study the proposal in \cite{Wyllard:2010} for cases involving general surface operators one would need to be able to  compute the corresponding  instanton partition functions. The main result of this paper is a conjecture which accomplishes this goal. Our conjecture is a generalisation to general partitions of $N$  of the results in \cite{Nekrasov:2002} and \cite{Feigin:2008} (which we reproduce for the cases $N=N$ and $N=1{+}\ldots{+}1$, respectively). We subject our conjecture to a variety of detailed tests and consistency checks. 

The analysis of  the consistency of our conjecture with the proposal in \cite{Wyllard:2010} is hampered by the fact that the $\cW$-algebras corresponding to a general partition of $N$ are not known in explicit form. However, for partitions of the type $N\!=\!1{+}\ldots{+}1{+}2$ the $\cW$-algebras are known as quasi-superconformal algebras and were written down by  Romans \cite{Romans:1990} (see also \cite{Fradkin:1992a} and section 5 in \cite{Kac:2003b}). The representation theory of the quasi-superconformal algebras was developed in \cite{Kac:2004}. Using these results we check that explicit $\cW$-algebra computations lead to results that are consistent with our proposed instanton partition functions.   

For other partitions of $N$ we do not know the proposed dual $\cW$-algebras explicitly. However, when $N\!=\!1+ (N{-}1)$  there is another known way a surface operator can arise. In the $6d$ language discussed above this type of surface operator arises from a $2d$ defect  spanning a $2d$ submanifold of $\RR^4$ and intersecting $C$ at a point. Such surface operators were first studied for the rank one case in  \cite{Alday:2009b} and correspond in the dual $2d$ CFT to the insertion of a certain degenerate field.  In \cite{Kozcaz:2010} it was shown that by combining the conjectures in \cite{Alday:2009a} and \cite{Alday:2009b} the instanton partition function with a surface operator of this type can be related to an $\SU(N){\times}\SU(N)$ quiver gauge theory {\it without} a surface operator but with certain restricitions  on the parameters of the quiver theory. This is a powerful method that works for any rank and leads to closed expressions using the results in \cite{Nekrasov:2002,Fucito:2004}. It was later realised \cite{Dimofte:2010,Taki:2010} that this type of argument corresponds to a geometric transition in the topological string language where the surface operator corresponds to a toric brane. Thus for $N\!=\!1+ (N{-}1)$ 
we have two types of surface operators: the ones that in the $6d$ language arise from $4d$ defects  and whose instanton partition function can be determined using the conjecture in this paper; and the surface operators that arise from $2d$ defects and whose instanton partition function can be determined from  a (constrained) $\SU(N){\times}\SU(N)$ theory using the results in \cite{Kozcaz:2010}.  

A priori it is not obvious that  the instanton partition functions arising from these two constructions should be related. But at least for theories where the gauge group has a single $\SU(N)$ factor computations in \cite{Awata:2010,Kozcaz:2010b} for $N=2$ and in \cite{Wyllard:2010} for $N=3$  indicate that the instanton partition functions do in fact agree. The analysis in  this paper for general $N$ lends further support to this idea (we find highly non-trivial agreement to several orders in perturbation theory and for some infinite subsets of terms).

The case $3\!=\!1{+}2$ studied in \cite{Wyllard:2010} belongs to both of the two classes mentioned above. The equivalence of the $\cW$-algebra and restricted quiver descriptions was checked in \cite{Wyllard:2010}. Taken together with the results in this paper we now have three different descriptions for this case which nicely illustrates all the interconnected conjectures.

In the next section we first recall some background material and introduce some terminology before stating our main conjecture which determines the instanton partition functions in $\SU(N)$ theories with a general surface operator. We also perform various general consistency checks. Then in section \ref{snfull} we treat the $N\!=\!1{+}\ldots{+}1{+}2$ cases and in section \ref{snfull} we discuss the $N\!=\!1+ (N{-}1)$ cases. We close with a brief summary and a discussion of some open problems.

\setcounter{equation}{0}
\section{Preliminaries and statement of the main conjecture}\label{sconj}

In this section, after discussing some relevant background material and setting up our notation, we write down a conjecture for  the instanton partition function in an $\cN=2$ $\SU(N)$ gauge theory with a general surface operator of the type discussed in the introduction. These surface operators  are classified by partitions of $N$ and for each such partition there is an associated instanton  partition function.  (In general there should also be perturbative contributions to the partition functions; such contributions will not be discussed in this paper.) 

The instanton partition functions should arise from integrals over (suitably regularised) moduli spaces of  instantons. In the presence of a surface operator the instantons are ``ramified instantons" that in a certain sense have both $4d$ and $2d$ contributions. We denote the regularised instanton  moduli spaces by $\widetilde{\cM}_{p; \vec{k}}$ where $p$ is a partition of $N$ and $\vec{k}=(k_1,\ldots,k_n)$ denote the instanton numbers. 
Schematically the instanton partition function is then given by
\be
Z_{\rm inst} = \sum_{\vec{k}} \prod_i y_i^{k_i} \int_{\widetilde{\cM}_{p; \vec{k}}} \om\,,
\ee
where $\om$ depends on the matter content of the theory. Various examples of such partition functions and moduli spaces have been considered in the literature (see e.g. \cite{Nekrasov:2002,Nakajima:2003a,Braverman:2004a,Feigin:2008,Negut:2008,Alday:2010,Awata:2010,Braverman:2010}). For the cases that have been proposed to describe gauge theories with surface operators (as first clearly stated in \cite{Alday:2010}), the integrals are well-defined as mathematical objects, but the physical interpretation is not yet completely clear. In this paper we follow \cite{Alday:2010,Awata:2010,Kozcaz:2010b,Wyllard:2010} and assume that the mathematical objects discussed in \cite{Braverman:2004a,Feigin:2008,Negut:2008,Braverman:2010} really describe instanton partition function for $\cN=2$ gauge theories with surface operators (more precisely the surface operators that in the $6d$ language arise from $4d$ defects as described in the introduction). So far there are no reasons to doubt this interpretation.

For a surface operator corresponding to a general partition of $N$ the regularised instanton moduli space should be an extension of the affine Laumon space considered in \cite{Feigin:2008} to the general parabolic case. In this case it appears hard to compute the integrals directly, but on general grounds one expects the integrals to localise to a set of isolated fixed points. In other words, it should be possible to write the instanton partition functions in the generic form 
\be
Z_{\rm inst} = \sum_{Y}  Z_{k_1,\ldots,k_{n}}(Y) \prod_i y_i^{k_i} \,,
\ee
where the sum is over a certain set of fixed points collectively labelled by  $Y$, the $y_i$ are the instanton expansion parameters and the $k_i$ are the instanton numbers (that are determined by $Y$). The expressions $Z_{k_1,\ldots,k_{n}}(Y)$ depend on the various parameters of the specific theory one considers such as the Coulomb parameters ($a$'s) and (if the theory includes matter fields) the masses ($m$'s) and finally also on two deformation parameters, $\ep_1$ and $\ep_2$ of the type used in  \cite{Nekrasov:2002}  (see also \cite{Moore:1997}).  
Furthermore, one expects that the  $Z_{k_1,\ldots,k_{n}}(Y)$ can be determined from a certain character. The character associated with a given fixed point takes the general form
\be \label{charexp}
\chi = \sum_i (\pm) e^{w_i} \,.
\ee
The contribution to the instanton partition function from the given fixed point  (denoted $Z_{k_1,\ldots,k_{n}}(Y) $ above) is then given by the product over the weights $w_i$, where  the weights coming from terms in (\ref{charexp}) with a minus sign contribute in the denominator and those arising from terms with a plus sign contribute in the numerator.

The basic building block used in computations is the character for a hypermultiplet  of mass $m$ transforming in the bifundamental  representation of $\SU(N){\times}\SU(N)$, which is of the general  (schematic) form
\be \label{bifchar}
\chi_{\rm bif} (a,\tia,Y,\tY,m) \,,
\ee
where $a$, $m$ refer to first $\SU(N)$ factor and $\tia$, $\tY$ refer to the second. 
From the expression (\ref{bifchar}) one can obtain the character for other representations of interest: The character for a matter multiplet of mass $m$ transforming in the adjoint representation of $\SU(N)$ is given by
\be \label{adj}
\chi_{\rm adj} (a,Y,m) = \chi_{\rm bif} (a,a,Y,Y,m)  \,,
\ee
the character of the gauge vector multiplet of $\SU(N)$ is obtained via
\be \label{vec}
\chi_{\rm vec} (a,Y) = -\chi_{\rm bif} (a,a,Y,Y,0)  \,,
\ee
and  the character for $N$ hypermultiplets transforming in the fundamental representation of the first (or second) $\SU(N)$ factor,  are  obtained via  
\bea \label{Nfunds}
\chi_{N \, {\rm funds}}(a,Y,\tilde{\mu}) = \chi_{\rm bif} (a,\tilde{\mu},Y,\emptyset,0) \,,\non \\
\chi_{N \, {\rm funds}}(\tia,\tY,\mu) = \chi_{\rm bif} (\mu,\tia,\emptyset,\tY,0)   \,.
\eea
Here $\mu$ and $\tilde{\mu}$ collectively denote the masses of the $N$ fundamentals, whereas $a$ and $\tia$ denote the $N$ Coulomb parameters. (Since the theory is $\SU(N)$ one should impose the restriction that the sum of the $a_i$ is zero and similarly for the $\tia_i$'s, but it is often convenient to leave this restriction implicit.) The total character  for an $\SU(N)$ quiver gauge theory is determined by summing the characters for all the constituent fields/representations of the theory.

Two cases where the above construction has been completed are known in the literature. The  first  case is the original Nekrasov construction \cite{Nekrasov:2002} (see also \cite{Flume:2002,Nakajima:2003a,Nekrasov:2003,Fucito:2004}) which is valid in the absence of surface operators. In this case the fixed points are labelled by a vector of $N$ Young tableaux $Y^I _\ell$, where $I=1,\ldots,N$ and $\ell$ label the columns of the Young tableau $Y^I$ (with $Y^I_1\ge Y^I_2\ge\cdots$). There is a single instanton expansion parameter, $y$, and the  instanton number $k$ is determined by 
\be \label{kN}
k= \sum_{I=1}^N \sum_{\ell\ge 1} Y_\ell^I. 
\ee

The second previously known case corresponds to the surface operator  labelled by the partition $N\!=\!1{+}\ldots{+}1$ (called a full surface operator in \cite{Alday:2010}). For this case the fixed points and the character were determined in~\cite{Feigin:2008}.   The fixed points are labelled by a periodic set of $N$ Young tableaux $Y^i_\ell$, with $Y^{i+N}=Y^i$. There are $N$ instanton expansion parameters $y_i$ (with $i=1,\ldots,N$) and the corresponding instanton numbers $k_i$ are determined by 
\be \label{kF}
k_i = \sum_{\ell\ge 1} Y_\ell^{i-\ell+1}. 
\ee

Our goal is to write down a conjecture for the character for an $\SU(N)$ theory in the presence of a general surface operator of the type discussed in the introduction. Such surface operators are classified by partitions of $N$. A partition of $N$ is a sum of integers $p_1+p_2+\ldots + p_n=N$, where $n$ is the length of the partition and the ordering is not important; in our conventions we always take $p_1\le p_2 \le \cdots \le p_n$. It will be convenient  to view the partitions as being periodic with period $n$, i.e.~$p_i \equiv p_{i+n}$. 

We now describe our construction. Just as in the previously studied cases, we assume that the fixed points are labelled by $N$ Young tableaux. These Young tableaux will be denoted $Y_\ell^{i, I_i}$, where $\ell \ge 1$ label the columns of  $Y^{i, I_i}$,  $i=1,\ldots,n$ and $I_i=1,\ldots p_i$.  To avoid cluttering the expressions we will usually drop the subscript on $I_i$ since the summation range should always be clear from the context. 
The Young tableaux are considered to be periodic in $i$ with period $n$ i.e.~$Y_\ell^{i, I_i}\equiv Y_\ell^{i+n, I_{i+n}}$. The $N$ Coulomb parameters will be denoted using a completely analogous notation as $a_i^{I_i}$. The Coulomb parameters are also considered to be periodic $a_i^{ I_i}\equiv a_{i+n}^{I_{i+n}}$. The instanton expansion parameters are denoted $y_i$ with $i=1,\ldots n$. 

In terms of these building blocks our conjectural expression for the character of a bifundamental hypermultiplet of mass $m$ in the $\SU(N){\times}\SU(N)$ gauge theory with a surface operator labelled by the partition $N=p_1+p_2+\ldots + p_n$ in each of the $\SU(N)$ factors takes the form
\bea \label{char}
&&  \!\!  \!\!  \!\!  \!\!  \!\!  \chi_{\rm bif} (a,\tia,Y,\tY,m) \,  = \, 
e^{-m} \sum_{i=1}^{n}  \sum_{I=1}^{p_i}  \sum_{\tell \ge 1} \sum_{\tI=1}^{p_{i-\tell} } 
e^{a_{i}^I - \tia_{i-\tell}^{\tI} } e^{\ep_2( \lfloor \frac{\tell-i}{n} \rfloor -  \lfloor -\frac{i}{n} \rfloor)}  \!\!\!\!\!  \sum_{\tilde{s}=1}^{\quad \tY^{i-\tell,\tI}_{\tell}}  \!\!\!  e^{\ep_1\tilde{s} } 
\non \\
&+&\!\!    (1{-}e^{\ep_1}) e^{-m} \sum_{i=1}^{n} \sum_{\ell\ge 1} \sum_{I=1}^{p_{i-\ell+1}}  \sum_{\tell \ge 1}\sum_{\tI=1}^{p_{i-\tell} } 
e^{a_{i-\ell+1}^I-\tia_{i-\tell}^{\tI}}e^{\ep_2( \lfloor \frac{\tell-i}{n} \rfloor -  \lfloor \frac{\ell-i-1}{n} \rfloor)} 
\!\!\!\!\!  \sum_{\tilde{s}=1}^{\quad \tY^{i-\tell,\tI}_{\tell}}  \!\!\!   e^{\ep_1\tilde{s}}  \!\!\!\!\!   \!\! \sum_{s=1}^{\quad Y^{i-\ell+1,I}_{\ell}}  \!\!\! \!\!\!  e^{-\ep_1s}
\non \\
&-&\! \!  (1{-}e^{\ep_1})  e^{-m}\sum_{i=1}^{n} \sum_{\ell\ge 1}\sum_{I=1}^{p_{i-\ell+1}}  \sum_{\tell \ge 1} \sum_{\tI=1}^{p_{i-\tell+1} }  
e^{a_{i-\ell+1}^I-\tia_{i-\tell+1}^{\tI}}e^{\ep_2( \lfloor \frac{\tell-i-1}{n} \rfloor -  \lfloor \frac{\ell-i-1}{n} \rfloor)}
 \!\!\!\!\!  \!\!  \sum_{\tilde{s}=1}^{\quad \tY^{i-\tell+1,\tI}_{\tell}}  \!\!\!   \!\!\! e^{\ep_1\tilde{s}}  \!\!\!\!\!   \!\! \sum_{s=1}^{\quad Y^{i-\ell+1,I}_{\ell}}  \!\!\!  \!\!\! e^{-\ep_1s} \non  \\
&+&\!\!   e^{\ep_1} e^{-m} \sum_{i=1}^{n} \sum_{\ell\ge 1} \sum_{I=1}^{p_{i-\ell+1}} \sum_{\tI=1}^{p_{i} } 
e^{a_{i-\ell+1}^I-\tia_{i}^{\tI}}e^{\ep_2( \lfloor -\frac{i}{n} \rfloor -  \lfloor \frac{\ell-i-1}{n} \rfloor)} \!\!\!\!\!  \sum_{s=1}^{\quad Y^{i-\ell+1,I}_{\ell}}  \!\!\!  e^{-\ep_1s } \,,
\eea
where $\lfloor x \rfloor$ denotes the largest integer smaller than or equal to $x$. In addition, the $n$ 
 instanton numbers are given by (note that $k_i \equiv k_{i+n}$)
\be \label{ki}
k_i = \sum_{\ell\ge 1} \sum_{I=1}^{p_{i-\ell+1}} Y_\ell^{i-\ell+1,I} \,.
\ee

As a first consistency check we now show that the above expressions reproduce the known results \cite{Nekrasov:2002} and \cite{Feigin:2008} when $N\!=\!N$ and $N\!=\!1{+}\ldots{+}1$, respectively.  
  
 For the $N\!=\!1{+}\ldots{+}1$ case, $n$ is equal to $N$ and the $I$'s and their sums can be removed (since in this case all the $p_i$ are 1 so all $I_i$ only take one value).  With these observations (\ref{ki}) immediately reduces to (\ref{kF}).  The above character (\ref{char}) also  reduces to the result obtained in \cite{Feigin:2008}. This is most easily seen using the form written in \cite{Kozcaz:2010b} and some obvious manipulations involving the geometric series
\be \label{geo}
\sum_{\tilde{s}=1}^{\tY} e^{ \ep_1 \tilde{s}} = e^{\ep_1}\frac{1-e^{\ep_1 \tY} }{1-e^{\ep_1} } \,.
\ee
  
For the $N\!=\!N$ case, $n$ is equal to $1$ and we can drop the $\lfloor \cdot \rfloor$  as well as the $i$'s and their sums (since $i$ only takes the value 1). The periodicity is 1 and can also be ignored. The expression (\ref{ki}) then directly reduces to (\ref{kN}).  Furthermore, it is easy to see that the character (\ref{char})  can be written
\bea
 \chi_{\rm bif} (a,\tia,Y,\tY,m) & = &  e^{-m} e^{\ep_1+\ep_2}\sum_{I=1}^N  \sum_{\tI=1}^N \sum_{\tell \ge 1} \sum_{\tilde{s}=1}^{\tY^I_{\tell} }e^{a_I - \tilde{\phi}_{\tI}} 
+ e^{-m} \sum_{\tI=1}^N\sum_{I=1}^N \sum_{\ell \ge 1} \sum_{s=1}^{Y^I_{\ell} } e^{\phi_I -\tia_{\tI}} 
\non \\
&-& e^{-m} (1-e^{\ep_1})(1-e^{\ep_2})  \sum_{I=1}^N \sum_{\ell \ge 1} \sum_{s=1}^{Y^I_{\ell} }    \sum_{\tI=1}^N \sum_{\tell \ge 1} \sum_{\tilde{s}=1}^{\tY^I_{\tell} }e^{\phi_I - \tilde{\phi}_{\tI}} \,,
\eea
where
\be
\phi_I = a_I - (\ell-1) \ep_1 - (s-1)\ep_2 \,, \qquad \quad \tilde{\phi}_{\tI} = {\tia}_{\tI} - (\tell-1) \ep_1 - (\tilde{s}-1)\ep_2 \,.
\ee
But this is precisely the usual form of the character \cite{Nekrasov:2002,Flume:2002,Fucito:2004} (with a particular sign convention for the $\ep_i$).
  
In the remainder of this  section we will discuss some further aspects of the above conjecture and perform some additional consistency checks for general partitions. 

In all known cases there is a connection between the dimension of the regularised instanton moduli space and the limit of the character when $a_i$ $\tia_i$, $\ep_1$ and $\ep_2$ are all taken to 0. In this limit we find that (\ref{char}) leads to the (conjectural) result
\be \label{dimM}
\dim \tilde{\cM}_{p,\vec{k}}  = \sum_{i=1}^{n} [ p_i k_{i-1} + p_i k_i ] \,.
\ee

In the expression (\ref{char}) terms with both positive and negative signs appear, but for e.g.~the  pure $\cN=2$ $\SU(N)$ theory, whose character is obtained using (\ref{vec}), one expects that there should be cancellations so that only one sign appears. 
We have not attempted to prove this is general but extensive perturbative computations lead us to conclude that  each power of $y_i$ in the instanton partition function for the  pure $\cN=2$ $\SU(N)$ theory with a general surface operator comes with a factor of the form 
\be \label{purestuff}
\prod_{a=1}^{ p_i+p_{i+1} } \frac{1}{w_a}\,.
\ee
When $N=2=1{+}1$ the result (\ref{purestuff}) agrees with the expression found in~\cite{Awata:2010}. 

The special case when $k_n=0$ is of particular interest. One may view this limit in various ways: In gauge theory language it is the limit when (in a certain sense) $4d$  instanton effects decouple, leaving only effects from ``$2d$ instantons"; in the CFT language it is the limit in which only the pieces independent of the worldsheet variable survive, and in which the $\cW$-algebra effectively  reduces to its finite version\footnote{For further information about  finite $\cW$-algebras see e.g.~\cite{Tjin:1992} and the references in \cite{Braverman:2010}.  
}.  For example in the case corresponding to $N\!=\!1{+}\ldots{+}1$ with $k_{N}=0$ the character obtained in \cite{Feigin:2008} reduces to the character in \cite{Negut:2008} (see e.g.~\cite{Kozcaz:2010b} for a discussion). In this example the $\cW$-algebra is the $\widehat{\sll}_N$ algebra and its finite version is the ordinary $\sll_N$ algebra. 

If we set $k_n=0$ in the general case then it is easy to see from (\ref{ki}) that only the following  finite set of components of the $Y^{i,I}$ may be non-zero: 
\bea
&& Y^{1,I} \,, \, \ldots  \,, Y^{1,I}_{n-2}  \,, \; Y^{1,I}_{n-1} \non \\
&& Y^{2,I} \,,  \, \ldots \,, Y^{2,I}_{n-2} \non \\
&&\;  \vdots  \qquad \oddots \\
&& Y^{n-1,I} \non 
\eea
This set of fixed points agrees with the result in \cite{Braverman:2010}, as do the expressions for the remaining non-zero $k_i$'s. Furthermore, when $k_n=0$ the dimension of the regularised moduli space (\ref{dimM}) reduces to $ \sum_{i=1}^{n-1} k_{i}( p_{i} +p_{i+1})$ which also agrees with the result in \cite{Braverman:2010} (in this case the relevant regularised moduli space is a so called parabolic Laumon space). These observations give further support to our conjecture. 

In the $k_{n}=0$ limit the character can be written in a simplified form. After some algebra we find
\bea \label{kn0}
\chi_{\rm bif} \left|_{k_N=0} \right. \! &=& \!  e^{-m}\sum_{i=1}^{n-1}\sum_{\ell=1}^{i+1} \sum_{I=1}^{p_{\ell}} \sum_{\tell=1}^{i} \sum_{\tI=1}^{p_{\tell}}  e^{a_\ell^I-\tia_{\tell}^{\tI} } \, \, \frac{ e^{ \ep_1( \tY_{i-\tell+1}^{\tell,\tI}-Y_{i-\ell+2}^{\ell,I} +1) }  -e^{\ep_1} } {e^{\ep_1}-1}  \non \\ 
&-&  \!  e^{-m}\sum_{i=1}^{n-1}\sum_{\ell=1}^{i} \sum_{I=1}^{p_{\ell}} \sum_{\tell=1}^{i} \sum_{\tI=1}^{p_{\tell}}  e^{a_\ell^I-\tia_{\tell}^{\tI} } \, \,  \frac{ e^{ \ep_1( \tY_{i-\tell+1}^{\tell,\tI}-Y_{i-\ell+1}^{\ell,I} +1) }  -e^{\ep_1} } {e^{\ep_1}-1}  \,.
\eea
This expression generalises the result in \cite{Negut:2008} to arbitrary partitions (note that (\ref{kn0}) only depends on $\ep_1$ and not on $\ep_2$). It was shown in \cite{Negut:2008} that the instanton partition function for the $\cN\!=\!2^*$ $\SU(N)$ theory (the theory with an adjoint hypermultiplet) with a $N\!=\!1{+}\ldots{+}1$ surface operator and $k_N=0$ equals (up to a prefactor) an eigenfunction of the quantum Calogero-Sutherland model. In a similar way, it is natural to expect that the partition function for the $\cN\!=\!2^*$ theory with a general surface operator  and $k_n=0$ is related to the eigenfunctions for some quantum integrable system.  
Other connections between eigenfunctions of quantum integrable systems and instanton partition functions (in the presence of surface operators) have been studied e.g.~in \cite{Nekrasov:2009,Awata:2009,Alday:2010,Maruyoshi:2010}). 

Finally, let us also mention a possible generalisation of our construction. The character for a  bifundamental hypermultiplet in (\ref{char}) involves the same surface operator in both factors of the gauge group. It may also be possible to extend this expression to the case when the surface operators in the two factors are different, but we will not attempt to do so here.

\setcounter{equation}{0}
\section{The case of $N=1{+}\ldots{+}1{+}2$ surface operators}\label{snfull}

In this section we test our conjectured instanton partition functions using the proposal in \cite{Wyllard:2010}. According to this proposal it should also be possible to compute the instanton partition functions for the $\cN=2$ $\SU(N)$ gauge theories (with surface operators of the type discussed in the introduction) from certain $\cW$-algebras. 

For the case of the $\SU(N)$ gauge theories with surface operators corresponding to partitions of the type $N\!=\!1{+}\ldots{+}1{+}2$  the corresponding $\cW$-algebras arising from the affine $\widehat{\sll}_N$ algebra by quantum Drinfeld-Sokolov reduction are known as quasi-superconformal algebras. These algebras were explicitly constructed by Romans \cite{Romans:1990} (see also \cite{Fradkin:1992a} and section 5 in \cite{Kac:2003b}), allowing us to make detailed checks of our instanton partition functions. 

We start by reviewing various aspects of the quasi-superconformal $\cW$-algebras and their representations.  These algebras are generated by the energy-momentum tensor $T(z)$, together with the   conformal dimension $3/2$  fields $G^{A}(z)$ and $\bar{G}_{A}(z)$,  the dimension 1 fields $J^A{}_B(z)$ forming an $\widehat{\sll}_{N-2}$ current algebra, and a dimension 1 scalar current $H(z)$.  In this section indices $A,B,\ldots$ take the values $2,\ldots,N{-}1$.  The mode expansions of the fields are the standard ones 
\bea \label{modes}
&&L(z) = \sum_n z^{-n-2} L_n \,, \qquad G^{A}(z) = \sum_n z^{-n-\frac{3}{2}} G^{A}_n \,, \qquad \bar{G}_A(z) = \sum_n z^{-n-\frac{3}{2}} \bar{G}_{n, A} \non \\ 
&& J^A{}_B(z) = \sum_n z^{-n-1} J^A_{n, B} \,, \qquad \qquad H(z) = \sum_n z^{-n-1} H_n  \,,
\eea
and the commutations relations of the modes   are
\bea \label{W32}
&&  \!\!  \!\!\!\!\! [L_n,L_m] = (n{-}m)L_{n+m} + \frac{c}{12} n(n^2-1)   \de_{n+m,0}  \,, \qquad [L_n,H_m] = -m\, H_{n+m}\,, \non \\
&&  \!\!   \!\!\!\!\! [L_n,J^A_{m, B}] = -m\, J^A_{n+m,B}\,, \quad [L_n,G_m^{A}] = (\frac{n}{2}{-}m)\, G^{A}_{n+m} \,, \quad [L_n,\bar{G}_{m, A}] = (\frac{n}{2}{-}m)\, \bar{G}_{n+m, A} \non 
\\ 
&& \!\!  \!\!\!\!\! [J^A_{n,B},J^C_{m, D}] = \de^C_B  \, J^A_{m+n,D} -  \de^A_D \, J^C_{m+n,B} + n\, (k{+}1)(\de^A_D\de^C_B - \frac{1}{N{-}2}\de^A_B \de^C_D) \, \de_{n+m,0}\,, \non
\\
&&  \!\!  \!\!\!\!\! [J^A_{n,B},G_m^{C}] = \de^C_B G_{n+m}^{A} - \frac{\de^A_B}{N{-}2}G_{n+m}^{C}  \,, 
\qquad  [J^A_{n,B}, \bar{G}_{m, C}] = -\de^A_C \bar{G}_{n+m, B} + \frac{\de^A_B}{N{-}2} \bar{G}_{n+m, C}  \,, \non \\
&&  \!\!   \!\!\!\!\!  [H_n,G_m^{A}] = \frac{G_{n+m}^{A}}{N{-}2}  \,, \qquad  [H_n,\bar{G}_{m, A}] = -\frac{\bar{G}_{n+m, A}}{N{-}2}  \,, \qquad  {}[H_n,H_m] = \frac{2k{+}N}{N(N{-}2)} \,n \, \de_{n+m,0} \,, 
\non \\ 
&&  \!\!   \!\!\!\!\!  {}[G^A_n, \bar{G}_{m, B}] = \de^A_B \bigg [ \frac{(k{+}1)(2k{+}N)}{2}(n^2{-}{\ts \frac{1}{4} }) \de_{n+m,0} - (k{+}N) L_{n+m}  +\frac{N}{2} (k{+}1) (n{-}m) H_{n+m}   \non \\
&& \qquad   + \, \frac{N(N{-}1)}{2}  \,\sum_\ell : \!H_{n+m-\ell}H_\ell \!: \bigg] + \frac{2k{+}N}{2} (n{-}m) J^A_{n+m,B} +  N \sum_\ell : \! H_{n+m-\ell}J^A_{\ell,B} \! :  \non \\
&&  \qquad  +\, \frac{1}{2} ( \de^A_C \de^D_B + \frac{1}{2} \de^A_B \de ^C_D)  \sum_\ell : \! \big( J^C_{n+m-\ell, E}J^E_{\ell,D} +  J^E_{n+m-\ell,D}J^C_{\ell, E}  \big)\! :  
\eea
where $k$ is a parameter and  $:\;:$ denotes the standard normal ordering of the modes
\be
: X_n Y_m: = \left\{ \ba{c}   X_n Y_m \qquad {\rm if} \qquad n \le  m \\     
Y_m X_n \qquad {\rm if} \qquad n > m\ea \right.
\ee
Finally, the central charge is  
\be
c=  \frac{-6k^2 + k\, (N^2 {-}5N{-}5) + N(N{-}4) }{k{+}N}\,.
\ee 
Note that our conventions differ slightly from the ones in \cite{Romans:1990}. When $N\!=\!3$ we recover the $\cW_3^{(2)}$ algebra \cite{Polyakov:1989}, that was studied in the present context in \cite{Wyllard:2010}. The quasi-superconformal algebras are similar to the $\cN=2$ superconformal algebra (which is the reason for their name), and have  both Ramond and Neveu-Schwarz sectors. We only consider the Ramond sector, where  $n$ in the mode-expansions  of $G^{A}(z)$  and $\bar{G}_{A}(z)$ in (\ref{modes}) are integers.  

In the Ramond sector, the properties of highest weight (or primary) states, $|\la \rb$, were worked out in \cite{Kac:2004}. The primary states are labelled by a vector, $\la$, in the root/weight space of $\sll_N$.  In the convention where the root/weight space is embedded in $\RR^{N}$ with unit vectors $u_L$ ($L=1,\ldots,N$)  we have 
\be
\la = \sum_{L=1}^{N} \la_L \, u_L \,,
\ee 
where it is implicitly understood that $\sum_{L=1}^N \la_L=0$. In this convention we find after translating the  results in \cite{Kac:2004} to our notation:
\bea \label{eigenvals}
&&  H_0 |\la \rb =  \left( - \frac{\la_1+\la_N}{N{-}2}  - \half \right) |\la \rb \,, \non \\ 
&& J^A_{0,A} |\la \rb = \left(  \la_{A}  + \frac{\la_1+\la_N}{N{-}2}\right)  |\la \rb \,, \qquad (A=2,\ldots,N{-}1) \,, \non \\
 && L_0 |\la \rb =  \left( \frac{ \lb \la, \la + 2 \rho - (k+N)\tha \rb}{2(k+N)} -\frac{N{-}2}{8} \right) |\la \rb 
 \,, 
  \eea
 where $\lb \cdot,\cdot\rb$ is the scalar product on the root space, $\rho= \half \sum_{L>M} u_L{-}u_M$ is the Weyl vector, and $\tha=u_N-u_1$ is the highest root. In addition to (\ref{eigenvals}) a highest weight state also satisfies
 \be \label{anni}
L_n |\la \rb =  G^{A}_{n} |\la \rb =  \bar{G}_{A, n} |\la \rb = H_n |\la \rb =  J^A_{n,B} |\la \rb =  0  \qquad (n=1,2,\ldots)\,,
\ee
and 
 \be \label{anni2}
G^A_{0} | \la \rb   = 0 \,, \qquad    J^A_{0, B}  |\la \rb =  0  \quad (A > B )\,.
\ee

The descendants of the primary  state $| \la \rb$ are obtained by acting with modes of the form  
\be
  G^{A}_{n} \,, \quad   \bar{G}_{n, A} \,, \quad H_{n} \,, \quad J^A_{n, B}  \,, \quad  L_{n}  \,,
\ee
where $n$ can be any negative integer. When $n=0$ we can also act with
\be
\bar{G}_{0, A}  \,, \qquad   J^A_{0, B}  \quad (A < B )\,.
\ee

We are only going to discuss the $\cW$-algebra duals to non-conformal theories. For these cases we need to introduce  Whittaker states (vectors).  These can be defined for the quasi-superconformal algebras in a way completely analogous to the construction in \cite{Braverman:2004a,Braverman:2010} (see also \cite{Wyllard:2010} and section 5 in \cite{Kozcaz:2010b}).  We denote the Whittaker state by $|\vec{x},z;\la \rb  \equiv |x_2,\ldots x_{N-1},z; \la \rb$   and demand that it  should satisfy
\bea \label{Wconds}
&&G_0^{2} |\vec{x},z; \la\rb = \sqrt{x_2} \,|\vec{x},z;\la \rb \,, \qquad \qquad \bar{G}_{1, N-1} |\vec{x},z;\la\rb = \sqrt{\frac{z}{x_{N-1}}}\, |\vec{x},z;\la \rb \,, \non  \\
&&J^3_{0,2} |\vec{x},z;\la\rb =\sqrt{\frac{x_3}{x_2}}\, |\vec{x},z; \la \rb \,, \quad \cdots \quad J^{N-1}_{0,N-2} |\vec{x},z; \la\rb =\sqrt{\frac{x_{N-1}}{x_{N-2}}}\, |\vec{x},z; \la \rb\,, 
\eea
where all other modes that annihilate $| \la \rb$ also annihilate $|\vec{x},z ; \la\rb$. When $N=3$ the second line in (\ref{Wconds}) is not relevant and we recover the expressions in \cite{Wyllard:2010}. 
The norm of the Whittaker state (sometimes known as an irregular conformal block) can be expressed in terms of certain (diagonal) components of the inverse of the matrix of inner products of descendants (i.e.~the Gram or Shapovalov matrix). The following set of descendants play a distinguished role in this construction
\be \label{des}
 |\vec{n}; \la\rb= (G^{N-1}_{-1})^{n_N} (J_0^{N-2}{}_{N-1})^{n_{N-1}} \cdots   (J^{2}_{0,3})^{n_{3}} (\bar{G}_{0,2})^{n_2} |  \la \rb \,.
\ee
The norm of the Whittaker vector can then be obtained via
\be \label{whit}
\lb \vec{x},z; \la |\vec{x},z; \la \rb =\sum_{n_2=0}^{\infty} \cdots  \!  \sum_{n_N=0}^{\infty} X_\la^{-1}(\vec{n};\vec{n}) \, x_2^{n_2} \left(\frac{x_3}{x_2}\right)^{n_3} \!\!\! \! \cdots   \left(\frac{x_{N-1}}{x_{N-2 }}\right)^{n_{N-1}} \!\!\left(\frac{z}{x_{N-1}}\right)^{n_N} \!\!.
\ee
where $X^{-1}_\la(\vec{n};\vec{n})$ denotes the diagonal component corresponding to (\ref{des}) of the inverse of the matrix of inner products of descendants. 
 
 From the proposal in \cite{Wyllard:2010} (see also \cite{Braverman:2010}) it follows that the  expression (\ref{whit}) should equal (possibly up  to a prefactor)  the instanton partition  function  for  the pure $\cN\,{=}\,2$ $\SU(N)$ theory with an $N=1{+}\ldots{+}1{+}2$ surface  operator insertion.

The subsets of terms in (\ref{whit}) that involve only one of the $N$ variables (i.e.~the terms with only one of the $n_L$ non-zero) can easily be computed. For the descendants that contribute to these terms the Gram matrix  is  diagonal and can trivially be inverted. As in \cite{Kozcaz:2010b} we find
\be
 \lb \la | (J^{A{+}1}_{0, \,A})^m  (J^{A}_{0, A{+}1})^m | \la \rb =   (-1)^m \,m! \, ( \la_{A}{-}\la_{A+1} )_{m} \quad \;  (A=2\ldots,N-2)\,.
\ee
 Using (\ref{W32}) together with  (\ref{eigenvals})-(\ref{anni2}) it can be shown that  
\bea \label{GG0}
 && \!\!\!\! m\, (\la_1 {-} \la_2{+} {\ts \frac{k}{2}} {+}{\ts \frac{N}{2}} {+}m{-}1)(\la_N {-} \la_2{-} {\ts \frac{k}{2}} {+}{\ts \frac{N{-}2}{2}} {+} m{-}1)\lb \la | (G^2_0)^{m-1} (\bar{G}_{0,2})^{m-1}|\la  \rb \non \\ \,    &\!\!\! =& \!\!\!  \lb \la|(G^2_0)^m (\bar{G}_{0,2})^m |\la \rb \, = \,  m! \, (\la_1- \la_2 +{\ts \frac{k}{2}} +{\ts \frac{N}{2}})_m (\la_N - \la_2 - {\ts \frac{k}{2}} +{\ts \frac{N{-}2}{2}})_m \,,
\eea
where $(X)_n= X (X+1) \cdots (X+n-1)$ is the usual Pochhammer symbol. 
Similarly, 
\be \label{GG1}
\lb \la|(\bar{G}_{1,N-1})^m (G^{N-1}_{-1})^m |\la \rb = (-1)^m\,  m! \, (\la_{N-1}{-}\la_1 -{\ts \frac{3k}{2}} -{\ts \frac{N{+}2}{2}})_m (\la_{N-1}{-}\la_N - {\ts \frac{k}{2}} -{\ts \frac{N}{2}})_m \,.
\ee
When $N=3$, (\ref{GG0}) and (\ref{GG1}) agree with the results in \cite{Wyllard:2010} after taking into account differences in conventions.

The contributions to (\ref{whit}) from the above classes of terms  are:
\bea \label{Wres}
&&\sum_{m=0}^{\infty}\frac{  1  }{  m! \, ( \la_{A}-\la_{A{+}1} )_{m} } \left(  - \frac{ x_{A+1}}{x_A} \right)^m  \qquad (A=2\ldots,N-2)  \,, \non \\
 &&\sum_{m=0}^{\infty}\frac{  1  }{   m! \,  (\la_1- \la_2 +{\ts \frac{k}{2}} +{\ts \frac{N}{2}})_m (\la_N - \la_2 - {\ts \frac{k}{2}} +{\ts \frac{N{-}2}{2}})_m }  \, x_2^m\,, \non \\
 &&\sum_{m=0}^{\infty}\frac{  1  }{  m! \,    (\la_{N-1}{-}\la_1 -{\ts \frac{3k}{2}} -{\ts \frac{N{+}2}{2}})_m (\la_{N-1}{-}\la_N - {\ts \frac{k}{2}} -{\ts \frac{N}{2}})_m }\left( - \frac{ z}{x_{N-1}} \right)^m .
\eea

Next we turn to the computation of the instanton partition function for the pure $\cN\,{=}\,2$ $\SU(N)$ theory with an $N=1{+}\ldots{+}1{+}2$ surface  operator insertion. In the notation of section \ref{sconj} we have $n=N-1$, $p_1=\ldots=p_{N-2}=1$ and $p_{N-1}=2$. From (\ref{char}) the terms with only one of the instanton expansion parameters $y_i$ non-zero can easily be computed. 
For these cases the instanton numbers involving the non-zero components of the Young tableaux are given by
\be
 k_j=Y^{j,1}_1  \quad (j=1,\ldots,N-2)\,,  \qquad k_{N-1}=  Y^{N-1,1}_{1} + Y^{N-1,2}_1 \equiv t_1 + t_2\,.
 \ee
 The corresponding characters become  (using the notation $a_j^1\equiv a_j$ ($j=1,\ldots,N{-}2$), $a_{N-1}^1 \equiv a_{N-1}$, and  $a_{N-1}^2 \equiv a_{N}$)
\bea
&&-(e^{a_{j+1}-a_j } +1)\sum_{s=1}^{k_j} e^{\ep_1 s}   \qquad \qquad  \qquad \qquad   \qquad \qquad(j=1,\ldots,N-3)\,, \non \\
&&-(e^{a_{j+1}-a_j }+ e^{a_{j+2}-a_j } +1)\sum_{s=1}^{k_j} e^{\ep_1 s}    \quad \qquad  \qquad \qquad \qquad(j=N-2) \,, \non \\
&&-(e^{a_{1}-a_{N-1}+\ep_2 } +1)\sum_{s=1}^{t_1} e^{\ep_1 s} -e^{a_{N-1}-a_{N}-t_2 } \sum_{s=1}^{t_1} e^{\ep_1 (s-t_1)} \\
&&-(e^{a_{1}-a_{N}+\ep_2 } +1)\sum_{s=1}^{t_2} e^{\ep_1 s} -e^{a_{N}-a_{N-1}-t_1 } \sum_{s=1}^{t_2} e^{\ep_1 (s-t_2)}    \qquad (j=N-1) \non 
\eea
These expressions translate into the following terms in the instanton partition function
\bea \label{res1}
&&\sum_{m=1}^{\infty} \frac{1}{m! \, (\frac{a_{i+1}-a_i}{\ep_1}+1)_m}\left( \frac{y_i}{(\ep_1)^2}\right)^m   \qquad \qquad \qquad \qquad  \qquad \; (i=1,\ldots,N-3) \,,\non \\
&&\sum_{m=1}^{\infty} \frac{1}{m! \,(\frac{a_{N-1}-a_{N-2}}{\ep_1}+1)_m (\frac{a_{N}-a_{N-2}}{\ep_1}+1)_m }\left( \frac{y_{N-2}}{(\ep_1)^3}\right)^m    \qquad (i=N-2) \,,
\eea
while for $i=N-1$ we find 
\bea \label{res2}
&&\sum_{m=1}^{\infty}  \sum_{t=0}^m   \frac{1}{t! \, (\frac{a_{1}-a_{N-1}+\ep_2}{\ep_1}+1)_t(\frac{a_{N}-a_{N-1}}{\ep_1}-m+t)_t }  \non  \\
&& \qquad \; \times \, \frac{1}{ (m-t)! \, (\frac{a_{1}-a_{N}+\ep_2}{\ep_1}+1)_{m-t}   (\frac{a_{N-1}-a_{N}-t}{\ep_1})_{m-t}  } \left( -\frac{y_{N-1}}{(\ep_1)^3}\right)^m 
\non \\
&=& \sum_{m=1}^{\infty} \frac{1}{m! \,(\frac{a_{1}-a_{N-1}+\ep_2 }{\ep_1}+1)_m (\frac{a_{1}-a_{N}+\ep_2 }{\ep_1}+1)_m }\left( \frac{y_{N-1}}{(\ep_1)^3}\right)^m  .
\eea
The results (\ref{res1}) and (\ref{res2}) for the instanton partition functions agree with the corresponding $\cW$-algebra expressions (\ref{Wres}) provided that we identify 
\be \label{map1}
\la_i - \half(N-2i+1) + \half(k{+}N)(\de_{i,1}-\de_{N,i}) = \frac{ a_{N-i}}{\ep_1}  \,, \quad \qquad k{+}N = - \frac{\ep_2}{\ep_1}
\ee
and
\be  \label{map2}
\frac{y_{i}}{(\ep_1)^2} =-  \frac{x_{N-i}}{x_{N-i-1}} \quad (1\le i \le N{-}3)\,, \quad \; \;   \frac{y_{N-2}}{  (\ep_1)^3}= x_2  \,,   \qquad \frac{y_{N-1}}{(\ep_1)^3}= -\frac{z}{ x_{N-1}}\,.
\ee
Note that the first relation in (\ref{map1}) can also be written 
\be
 \frac{a}{\ep_1} =  \la +\rho  -\half (k{+}N)\tha \qquad (\mathrm{where} \quad a\equiv \sum_{i=1}^N a_{N-i} \, u_i) \,.
\ee

Just as in the previous work \cite{Kozcaz:2010b,Wyllard:2010} it is also possible to study higher-order corrections. For instance, the computations of  $y_i^m y_{i+1}$ terms with $i=1,\ldots, N{-}4$  are completely analogous to the computations in \cite{Kozcaz:2010b} and the computation of terms of the form $y_{N-2}^m y_{N-1}$ is completely analogous to the computation in \cite{Wyllard:2010}. Terms of the form $y_i^n y_j$ with $j\neq i, i\pm1$ can also be seen to have the right general structure. We will not give any further details here. 

We should also mention that most of the terms we discssed above (with the exception of those that in the $\cW$-algebra language depend on $z$) are covered by the proof in \cite{Braverman:2010}, i.e.~for these terms it has been proven that they agree with the integrals over instanton moduli space. Thus for these terms our computation should be viewed as a check that our combinatorial expressions really equal the values of the integrals. 

In this section we focused on the quantities that on the gauge theory side correspond to the (non-conformal) pure $\SU(N)$ theories. It should also be possible to consider conformal  $\SU(N)$ gauge theories, but  one would need to overcome the technical difficulties discussed in \cite{Wyllard:2010}.

\setcounter{equation}{0}
\section{The case of $N=1{+}(N{-}1)$ surface operators}\label{ssimp}

In the previous section we discussed the dual $\cW$-algebra description for the class of surface operators corresponding to the partitions $N\!=\!1{+}\ldots{+}1{+}2$. Unfortunately, a similar analysis is not possible in the general case since we do not know the relevant $\cW$-algebras in explicit form, but as mentioned in the introduction when $N\!=\!1{+}(N{-}1)$ there is a construction involving an alternative type of surface operator that we can compare our instanton partition functions to. 

This alternative construction arises since from the $6d$ perspective it is also possible to obtain  surface operators using $2d$ defects spanning a submanifold inside $\RR^4$ and intersecting $C$ at a point (see \cite{Alday:2010} for a discussion). This construction was pioneered in \cite{Alday:2009a}  and leads to a surface operator corresponding to the partition\footnote{So far this is the only case that has been treated using this setup. It is unclear to us if more general surface operators can also be treated using this approach.}  $N=(N{-}1)+1$. Thus for such partitions there are two types of surface operators, those that (in the $6d$ language) arise from $4d$ defects and those that arise from $2d$ defects. These are a priori distinct objects, but as mentioned in the introduction there are indications that they lead to identical instanton partition functions, at least for some theories. In particular, we expect agreement for the theories  where the gauge group is $\SU(N)$. This expectation is based on the computations in \cite{Awata:2010,Kozcaz:2010b} for $N=2$ and in \cite{Wyllard:2010} for $N=3$. In this section we extend the analysis in \cite{Awata:2010,Kozcaz:2010b,Wyllard:2010} to general $N$. 

To proceed we first recall the main facts about the construction of surface operators using the $2d$ defects. It was argued in \cite{Alday:2009b} that such a surface operator in an $\SU(2)$ gauge theory  has a dual description in the Liouville theory in terms of  the insertion of a certain degenerate field localised at the point where the defect intersects $C$.  In a later development  \cite{Kozcaz:2010} it was shown  that by combining the conjectures in \cite{Alday:2009a} and \cite{Alday:2009b}  one can obtain (conjectural) closed expressions for the gauge theory instanton partition function in $\SU(N)$ theories with this type of  surface operator (the method works for both conformal and non-conformal theories).  Further aspects  have been studied e.g.~in \cite{Gaiotto:2009c,Kozcaz:2010,Dimofte:2010,Maruyoshi:2010,Taki:2010,Marshakov:2010}. In particular, there are interesting topological string constructions but these will not be discussed here.

For simplicity we only discuss the method in \cite{Kozcaz:2010} for the pure $\SU(N)$ theory, but the extension to theories with matter is straightforward (see section 6 of  \cite{Kozcaz:2010} for more details). The starting point is the instanton partition function for the $\SU(N){\times}\SU(N)$ theory with one bifundamental hypermultiplet. This partition function can be written 
\bea \label{Z2quiver}
&& \sum_{Y, \tY } y^{k} \, \tilde{y}^{\tilde{k}}  \prod_{I,J=1}^{N} \prod_{s\in Y^I} \frac{ [m-E(a_I-\tia_J,Y^I,\tY^J,s)]  }{E(a^I-a^J,Y^I,Y^J,s)\, [E(a_I-a_J,Y^I,Y^J,s) +\ep] } \non \\ 
&& \qquad \qquad \quad \, \times \prod_{t\in \tY^I} \frac{[m+E(\tia_I-a_J,\tY^I,Y^J,t)+\ep] }{E(\tia_I-\tia_J,\tY^I,\tY^J,t)\, [E(\tia_I-\tia_J,\tY^I,\tY^J,t) +\ep] } 
\,,
\eea
where $\ep=\ep_1+\ep_2$, the $Y$ sum is over all  $N$-dimensional vectors of Young tableaux $(Y^1,Y^2,\ldots,Y^{N})$, the instanton number $k$ is as in (\ref{kN}), the $a_I$ are the Coulomb  moduli for the first $\SU(N)$ factor, and $\tY$, $\tilde{k}$, and $\tia_I$ denote the corresponding quantities referring to  the second factor. Furthermore (in our conventions)
\be \label{rest}
E(x,Y^I,W^J,s) = x +\ep_1 L_{W^{J}}(s) - \ep_2(A_{Y^I}(s)+1)\,,
\ee
 where $s=(j,\ell)$ and $j$ refers to the vertical position and $\ell$ to the horizontal position of a box in the Young tableau $Y^I$. In (\ref{rest}), $L_{Y^J} = (Y^J)^T_{j}-\ell$ and $A_{Y^I} = Y^I_{\ell} - j$, where $Y^I_{\ell}$ is the height of the $\ell$th column of $Y^I$, and $(Y^J)^T_{j}$ is the height of the $j$th column of the transpose of $Y^J$ (i.e.~the length of the $j$th row of $Y^J$). 

The next step is to impose the restrictions 
\be
 m=\frac{ \ep_2 }{ N }\,, \qquad \qquad  \tia_I = a_I + \ep_2 \,  \La_1 \,,
 \ee
 where $\La_1 = \frac{1}{N}(N-1,-1\ldots,-1)$ is the first fundamental weight of the $A_{N-1}$ Lie algebra.   These conditions are simply the translation of the degenerate field and degenerate fusion results  from the $A_{N-1}$ Toda theory/$\cW_N$-algebra to the gauge theory variables using the AGT relation.   

The expression (\ref{Z2quiver}) with the restrictions (\ref{rest}) gives a closed  expression for the partition function for the pure $\SU(N)$ theory with a $N\!=\!1{+}(N{-}1)$ surface operator arising from a $2d$ defect. In particular, when $k=0$ (i.e.~$Y$ is empty) it is easy to see that only $\tY^1$ can be non-zero and furthermore can have boxes only in the first column otherwise the expression (\ref{Z2quiver}) with (\ref{rest})  vanishes (similar arguments were  used in \cite{Mironov:2009a}). This implies that the partition function reduces to 
\be \label{yn1}
\sum_{ \tilde{k}=1}^{\infty}  \frac{1}{ \tilde{k}! \, \prod_{I=1}^{N-1} ( \frac{a_{I+1}}{\ep_1}-\frac{a_{1}}{\ep_1} +1 )_{\tilde{k} } }    \left( \! \frac{\tilde{y}}{(\ep_1)^N}\!  \right)^{\tilde{k} }.
\ee
This result can also be understood from the Toda CFT point of view using the results in \cite{Fateev:2005} in the confluent limit where the ${}_N F_{N-1}$ hypergeometric function is reduced to ${}_0F_{N-1}$ (see e.g.~\cite{Kozcaz:2010b,Wyllard:2010} for similar discussions).   

Next we consider the pure $\SU(N)$ theory with a $N\!=\!1{+}(N{-}1)$ surface operator arising from a $4d$ defect using our conjectured instanton partition function (\ref{char}). In the special case $k_1\neq 0$ but $k_2=0$ it follows from (\ref{ki}) that we necessarily have $Y^{2,I}_\ell=0$ and $k_1 = Y^{1,1}_1$ with all other components of $Y^{1,I}_\ell$ vanishing. Then (\ref{char}) implies that for the  pure $\cN=2$ $\SU(N)$ theory the character corresponding to the $y_1^k$ term in the instanton expansion becomes (here we have relabelled the Coulomb parameters $a_1^1\rar a_1$, $a_2^I\rar a_{I+1}$)
\be
\left[ 1+ \sum_{I=1}^{N-1} e^{a_{I+1}-a_1} \right]\sum_{s=1}^{k} e^{\ep_1 s}  \,.
\ee
This result leads to the following terms in the instanton partition function for the pure $\cN=2$ $\SU(N)$ theory
\be  \label{yn2}
\sum_{k_1=1}^{\infty} \frac{1}{ (k_1)! \prod_{I=1}^{N-1} (\frac{a_{I+1}}{\ep_1}-\frac{a_{1}}{\ep_1}+1 )_{k_1}  }  \left( \! \frac{y_1}{(\ep_1)^N} \! \right)^{k_1} ,
\ee
which agrees perfectly with (\ref{yn1}) if we identify $(\tilde{y},y)=(y_1,y_2)$ and $(\tilde{k},k)=(k_1,k_2)$. 

Similarly one can also analyse the terms with $k_1=0$ ($\tilde{k}=0$). In both descriptions it can be shown that the partition function becomes
\be 
\sum_{k_2=1}^{\infty}   \frac{1}{k! \prod_{I=1}^{N-1} (\frac{a_{1}}{\ep_1}-\frac{a_{I+1}}{\ep_1}+\frac{\ep_2}{\ep_1}+1 )_{k_2} }  \left( \! -\frac{y_2}{(\ep_1)^N} \! \right)^{k_2} .
\ee

In addition to the above two infinite sets of terms we have also performed several perturbative checks for low ranks,  finding highly non-trivial agreement between the two descriptions. For instance for $N=4$ we checked the agreement of the two constructions up to total instanton number $k_1+k_2=4$. 

Note that although the quiver expression (\ref{Z2quiver}) with the restrictions (\ref{rest}) also arises from a localisation problem, it  is not true that the agreement with the result in section \ref{sconj} is  fixed-point by fixed-point; in other words it is  only after summing up all terms with given instanton numbers $k_1$, $k_2$ that the two expressions agree. Nevertheless, since we have explicit formul\ae{} in both cases it is possible that a proof can be found but it will not be attempted here.

\setcounter{equation}{0} 
\section{Discussion} \label{sdisc}

In this paper three different interconnected conjectures have entered the analysis. The first conjecture is the explicit formula for the instanton partition function in an $\cN=2$ $\SU(N)$ gauge theory with a general surface operator encoded in (\ref{char}).  The second conjecture is the proposal in \cite{Wyllard:2010} that instanton partition functions in  $\SU(N)$ gauge theories with a general surface operator should also be computable from the corresponding $\cW$-algebra. The final conjecture is the assumption that in an $\cN=2$ $\SU(N)$ gauge theory, the  $N\!=\!1+(N{-}1)$ $\SU(N)$ surface operators  which in the $6d$ language are constructed from $4d$ or $2d$ defects lead to identical instanton partition functions.   

We have found striking and highly non-trivial agreements between these conjectures,  nevertheless we should stress that none of these conjectures have been proven (with the exception of the results in  \cite{Braverman:2010}) nor are their physical origin clearly understood. 

It is important to study the general instanton partition functions further and clarify their meaning. The special case $k_n=0$  may prove a useful testing ground.  
For the cases without surface operators one does not just have the formulation in terms of the combinatorial Nekrasov expressions, but there is also a formulation in terms of so called LMNS contour integrals \cite{Moore:1997} (in the recent developments, this formulation was for instance useful in \cite{Fateev:2009}). One could ask if a similar formulation is also possible for the cases with surface operators. 

The classes of partitions/surface operators that we discussed in this paper are somewhat special. In the equivalent language of nilpotent orbits, we discssed the zero orbit ($1{+}\ldots{+}1$), the minimal orbit ($1{+}\ldots{+}1{+}2$), the subregular orbit  ($1+(N{-}1)$) and the principal orbit ($N$). In the Hasse diagram these orbits are the two entries on the top and at the bottom. It would be interesting to also study  some intermediate cases.
 
 It is known that the instanton partition functions for the  $N\!=\!1{+}\ldots{+}1$ case satisfy certain differential equations. For the pure case this was established in \cite{Braverman:2004a}; see also \cite{Awata:2009}. For the case with matter the differential equation was proposed in the recent paper \cite{Yamada:2010}. It seems plausible to expect that also the partition functions for other surface operators satisfy interesting differential equations.

\section*{Acknowledgements}

I would like to thank Can Koz\c{c}az,  Sara Pasquetti and Filippo Passerini for discussions.



\providecommand{\href}[2]{#2}\begingroup\raggedright\endgroup

\end{document}